\documentclass[twocolumn,showpacs,prb]{revtex4}

\usepackage{graphicx}%
\usepackage{dcolumn}
\usepackage{amsmath}

\makeatletter
\def\btt#1{\texttt{\@backslashchar#1}}%
\DeclareRobustCommand\bblash{\btt{\@backslashchar}}%
\makeatother

\begin{document}

\title[Short Title]{Split of zero-bias conductance peak in normal-metal / 
$\boldsymbol{d}$ wave superconductor 
junctions}

\author{Yasuhiro Asano}
\email{asano@eng.hokudai.ac.jp}
\affiliation{
Department of Applied Physics, Hokkaido University, Sapporo 060-8628, Japan}%

\author{Yukio Tanaka}
\affiliation{
Department of Applied Physics, Nagoya University, Nagoya 464-8603, Japan}%

\author{Satoshi Kashiwaya}
\affiliation{
National Institute of Advanced Industrial Science and Technology, 
Tsukuba, 305-8568, Japan}%

\date{\today}

\begin{abstract}
Effects of impurity scatterings on the conductance in normal-metal / $d$ wave superconductor 
junctions are discussed by using the single-site approximation.
So far, the split of the zero-bias conductance peak has been believed to be an evidence
of the broken time reversal symmetry states at the surface of high-$T_c$ superconductors.
In this paper, however, it is shown that the impurity scattering near the interface also 
causes the split of the zero-bias conductance peak.
Typical conductance spectra observed in experiments at finite temperatures and under external 
magnetic fields are explained well by the present theory.
\end{abstract}

\pacs{74.81.-g, 74.25.Fy, 74.50.+r}
\maketitle

\section{introduction}
The zero-energy state (ZES)~\cite{hu} formed at surfaces of superconductors is 
a consequence of the unconventional symmetry of Cooper
pairs. Since the ZES appears just on the Fermi energy, 
it drastically affects transport properties through the interface of junctions 
consist of unconventional superconductors.~\cite{RPP} 
For instance in normal-metal / high-$T_c$ superconductor junctions, a large peak
is observed in the differential conductance at the zero 
bias voltage.~\cite{tanaka0,kashiwaya,Alff,Wang,wei,iguchi,geerk,mao}
The ZES is also responsible for the low-temperature anomaly of the Josephson
current between the two unconventional 
superconductors.~\cite{barash,tanaka1,tanaka2,tanaka3,T1,asano,arie,ilichev1,ilichev2,shirai,testa}

An electron incident into a normal-metal / superconductor (NS) interface suffers 
the Andreev reflection~\cite{andreev} by the pair potential in the superconductor. 
As a result, a hole traces back 
the original propagation path of the incident electron. 
This is called the retro property
of a quasiparticle which supports the formation of the ZES.
Strictly speaking, the electron-hole pairs just on the Fermi energy 
hold the retro property in the presence of the time reversal symmetry (TRS).
Thus the ZES is sensitive to the TRS of the system.
Actually, the zero-bias conductance peak (ZBCP) in NS junctions splits into two peaks 
under magnetic fields.~\cite{fogelstrom,YT021,YT022,YT023}
The peak splitting is also discussed~\cite{kashi95,matsumoto2,TJ1,TJ2,Tanuma2001,lubimova,kitaura} 
when the broken time reversal symmetry state (BTRSS) is formed at the interface.
Theoretical studies showed that such BTRSS's are characterized by the 
$s+id_{xy}$~\cite{matsumoto2} or $d_{xy}+id_{x^2-y^2}$~\cite{laughlin} wave pairing 
symmetry. 
Experimental results, however, are still controversial.
Some experiments reported the split of the ZBCP at the 
zero magnetic field,~\cite{covington,biswas,dagan,sharoni,kohen,greene,aprili2,krupke}
other did not observe the splitting.~\cite{Ekin,Alff,wei,iguchi,Sawa1,Aubin} 
The ZBCP is also sensitive to the exchange potential in ferromagnets attaching
to unconventional superconductors.~\cite{H2,Y2}

In previous papers, we numerically showed that random potentials at the 
NS interface cause the split of the ZBCP at the zero magnetic field by 
using the recursive Green function method.~\cite{asano2,asano3}
We also showed that the splitting due to the impurity scattering can 
be seen more clearly 
when realistic electronic structures of high-$T_c$ materials are 
taken into account.~\cite{asano4}
Unfortunately, we could not make clear a mechanism of splitting.
Our conclusion, however, contradicts to those of a number of 
theories~\cite{barash2,golubov,poenicke,yamada,tanaka5,luck}
based on the quasiclassical Green function 
method.~\cite{eilenberger,larkin,zaitsev,shelankov,bruder} 
The drastic suppression of the ZBCP by the interfacial randomness is 
the common conclusion of all the theories. The theories of the 
quasiclassical Green function method, however, concluded 
that the random potentials do not split the ZBCP.
Thus this issue has not been fixed yet.
There are mainly two reasons for 
the disagreement in the two theoretical approaches (i.e., the recursive Green function
 method and the quasiclassical Green function method).
One is the treatment of the random potentials, the other is the effects of the 
rapidly oscillating wave functions on the conductance.
In our simulations, we calculate the conductance without 
any approximation to the random potentials and the wave functions; this is an 
advantage of the recursive Green function method.~\cite{lee,asano3}

In this paper, we discuss effects of the impurity scattering on the conductance
in normal-metal/$d$ wave superconductor junctions
by using the Lippmann-Schwinger equation. We assume that impurities 
are near the NS interface on the superconductor side. 
The differential conductance is analytically calculated 
 within the single-site approximation
based on the conductance formula.~\cite{blonder,takane}
The split of the ZBCP due to the impurity scattering is the main 
conclusion of this paper.
The impurity scattering affects the conductance in two ways: (i)
drastically suppressing the conductance around the zero bias voltage and (ii)
making the conductance peak wider.
The split of the ZBCP is a consequence of the interplay between the two effects.
In the present theory, we successfully
explain typical conductance shapes observed in several experiments.
We also show that the splitting peaks are merged into a single conductance peak 
for sufficiently high temperatures and that the peak splitting width increases with
increasing external magnetic fields.

This paper is organized as follows. In Sec.~II, we derive
the reflection coefficients in NS junctions within the single-site
approximation based on the Lippmann-Schwinger equation.
The split of the ZBCP is discussed in Sec.~III.
In Sec.~IV, calculated results are compared with experiments and another theories.
In Sec.~V, we summarize this paper.

\section{Lippmann-Schwinger equation}
\begin{figure}[htbp]
\begin{center}
\includegraphics[width=8.0cm]{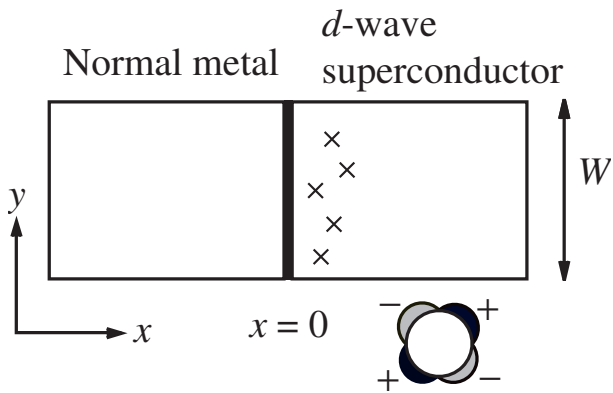}
\end{center}
\caption{
The normal-metal / $d$-wave superconductor junction is schematically illustrated.
The crosses represent impurities.}
\label{system}
\end{figure}

Let us consider two-dimensional NS junctions as shown in Fig.~\ref{system},
where normal metals ($x<0$) and $d$ wave superconductors ($x>0$) 
are separated by the potential barrier $V_B(\boldsymbol{r})=V_{0} \delta(x)$.
We assume the periodic boundary condition in the $y$ direction and the 
width of the junction is $W$.
The $a$ axis of high-$T_c$ superconductors is oriented by the 45 degrees from
the interface normal. The pair potential of a high-$T_c$ superconductor is described by
\begin{equation}
\Delta_{\boldsymbol{k}} = 2 \Delta_0 \bar{k}_x \bar{k}_y,
\end{equation}
in the momentum space, where $\Delta_0$ is the amplitude of the pair potential at the 
zero temperature, $\bar{k}_x = \cos\gamma={k}_x /k_F$ and $\bar{k}_y = \sin\gamma
={k}_y/k_F$ 
are the normalized wave number on the Fermi surface in the $x$ and the $y$ direction, respectively.
The Fermi wave number $k_F$ satisfies $ \hbar^2k_F^2/2m=\mu_F$, where $\mu_F$ is the Fermi energy. 
The schematic figure of the pair potential is shown in Fig.~\ref{system}.
The NS junctions are described by the Bogoliubov-de Gennes equation,~\cite{degennes}
\begin{align}
\int d{\boldsymbol{r}'} &\left( \begin{array}{cc}
\delta(\boldsymbol{r}-\boldsymbol{r}')h_0(\boldsymbol{r}') & \Delta(\boldsymbol{r},\boldsymbol{r}') \\
\Delta^\ast(\boldsymbol{r},\boldsymbol{r}') & -\delta(\boldsymbol{r}-\boldsymbol{r}')
h_0(\boldsymbol{r}')\end{array}\right) \nonumber \\
& \times\left( \begin{array}{c} u(\boldsymbol{r}') \\  v(\boldsymbol{r}') \end{array} \right) 
 = E  \left( \begin{array}{c} u(\boldsymbol{r}) \\  v(\boldsymbol{r}) \end{array} \right),
\end{align}
\begin{align}
h_0(\boldsymbol{r}) =& -\frac{\hbar^2 \nabla^2}{2m} + V_{pot}(\boldsymbol{r}) - \mu_F, \\
V_{pot}(\boldsymbol{r}) =& V_B(\boldsymbol{r}) + V_I(\boldsymbol{r}),\\
\Delta(\boldsymbol{R_c},\boldsymbol{r}_r)= &\left\{\begin{array}{ccc}  
\frac{1}{V_{vol}}\sum_{\boldsymbol{k}} \Delta_{\boldsymbol{k}} e^{i\boldsymbol{k}\cdot \boldsymbol{r}_r} & : &
 X_c > 0\\
0  & : &  X_c < 0 
\end{array}\right.,
\end{align}
where $\boldsymbol{R}_c=(X_c,Y_c)=(\boldsymbol{r}+\boldsymbol{r}')/2$ and 
$\boldsymbol{r}_r=\boldsymbol{r}-\boldsymbol{r}'$. 
Throughout this paper, we neglect the spatial dependence of 
the pair potential near the junction interface.
This is a reasonable approximation when we consider the conductance
around the zero bias voltage.~\cite{tanaka4}
The spatial dependence of 
the pair potential should be determined in a self-consistent way
when we discussed the conductance far from the zero bias such as $eV \sim \Delta_0$.
Here $V$ is the bias voltage applied to junctions.

We consider impurities near the interface on the superconductor side as
indicated by crosses in Fig.~\ref{system}. 
The potential of impurities is given by
\begin{equation}
V_I(\boldsymbol{r}) = V_i \sum_{j=1}^{N_i} \delta(\boldsymbol{r}-\boldsymbol{r}_j),
\end{equation}
where $N_i$ is the number of impurities.
In the absence of impurities, the transmission and the reflection coefficients
are calculated from boundary conditions of wave functions at the junction interface 
as shown in Appendix~A.
By using these coefficients,
four retarded Green functions are obtained as shown in Appendix~B.
The normal conductance of the junction is given by
\begin{align}
G_N =&\frac{2e^2}{h} N_c T_B,\\
T_B =& \int_0^{\pi/2} \!\!\! d\gamma \, \frac{ \cos^3\gamma}{z_0^2+\cos^2\gamma},
\end{align}
where 
 $T_B$ is the transmission probability of the junction, $N_c=2W/\lambda_F$ is the
number of the propagating channels on the Fermi surface, $\lambda_F=2\pi/k_F$ is the
Fermi wave length and 
$z_0=mV_{0}/(\hbar^2 k_F)$ represents the strength of the potential barrier
at the NS interface. In the limit of $z_0^2 \gg 1$, $T_B$ 
 is proportional to $1/z_0^2$.

Effects of impurities on the wave functions are taken into account by 
using the Lippmann-Schwinger equation,
\begin{align}
{\psi}^{(l)}(\boldsymbol{r}) =& {\psi}^{(l)}_0(\boldsymbol{r})
+ \int\!\! d\boldsymbol{r}' \hat{G}_0(\boldsymbol{r},\boldsymbol{r}')\;V_I(\boldsymbol{r}') 
\;\hat{\sigma}_3 \;
{\psi}^{(l)}(\boldsymbol{r}'),\\
=& {\psi}^{(l)}_0(\boldsymbol{r})
+ \sum_{j=1}^{N_i}\hat{G}_0(\boldsymbol{r},\boldsymbol{r}_j)\;V_i \;\hat{\sigma}_3 \;
{\psi}^{(l)}(\boldsymbol{r}_j),\label{ls1}
\end{align}
where $l$ indicates a propagating channel characterized by the transverse
wave number $k_y^{(l)}$.
Here ${\psi}^{(l)}_0(\boldsymbol{r})$ is the wave function in which
an electronlike quasiparticle with $k_y^{(l)}$ is incident into the clean NS 
interface from normal metals and is described as
\begin{align}
{\psi}^{(l)}_0(\boldsymbol{r})=& \chi_l(y)
\left[
\left(\begin{array}{c} 1\\0\end{array} \right) e^{iq_l^+x} +  
\left(\begin{array}{c} 0\\1\end{array} \right) e^{iq_l^-x} \; r^{he}_{NN}(l) 
\right.\nonumber\\
&+ \left.
\left(\begin{array}{c} 1\\0\end{array} \right) e^{-iq_l^+x} \; r^{ee}_{NN}(l)\right],\\
\chi_l(y)=&\frac{e^{ik_y^{(l)}y}}{\sqrt{W}},
\end{align}
for $x<0$, where $q^\pm_l=  \sqrt{ k_l^2 \pm k_F^2 {E}/{\mu_F}}$ is the wave number of a
quasiparticle in normal metals and $k_l^2 + {k_y^{(l)}}^2 = k_F^2$.
For $x>0$, the wave function in clean junctions is given by 
\begin{align}
{\psi}^{(l)}_0(\boldsymbol{r})=& \chi_l(y)
\hat{\Phi} \left[ 
\left(\begin{array}{c} u_l\\v_l \end{array} \right) e^{ik^+_lx}\;  
t^{ee}_{SN}(l) \right.\nonumber\\
&+ \left. \left(\begin{array}{c} -v_l \\ u_l \end{array} \right) e^{-ik^-_lx}\;  
t^{he}_{SN}(l) \right],\\
u_l  =& \sqrt{ \frac{E + \Omega_l}{2E}},\\
v_l =& \textrm{sgn}({k}_y^{(l)}) \sqrt{ \frac{E -\Omega_l}{2E}},\\
\hat{\Phi}=&\left( \begin{array}{cc} e^{i\varphi} & 0 \\
0 & e^{-i\varphi}\end{array}\right),
\end{align}
where $\varphi$ is a macroscopic phase of a superconductor, $k^\pm_l = \left( k_l^2 \pm k_F^2{\Omega_l}/{\mu_F}\right)^{1/2}$ is the wave number 
of a quasiparticle in superconductors,
$\Omega_l=\sqrt{ E^2 - \Delta_l^2}$ and
$\Delta_l=2\Delta_0 \bar{k}_l\bar{k}_y^{(l)}$.
The wave function at an impurity ${\psi}^{(l)}(\boldsymbol{r}_{j'})$ can be obtaind by 
$\boldsymbol{r}\to \boldsymbol{r}_{j'}$ in Eq.~(\ref{ls1})
\begin{align}
{\psi}^{(l)}_0(\boldsymbol{r}_{j'}) =&\! \sum_{j=1}^{N_i}\!\! \left[ \hat{\sigma}_0
\delta_{j,j'}
- \hat{G}_0^{SS}(\boldsymbol{r}_{j'},\boldsymbol{r}_j)\;V_i \;\hat{\sigma}_3\right]
{\psi}^{(l)}(\boldsymbol{r}_j). \label{psi-1}
\end{align}
It is possible to calculate the exact conductance if we 
obtain ${\psi}^{(l)}(\boldsymbol{r}_j)$ for all impurities by solving Eq.~(\ref{psi-1}). 
Actually it was confirmed that the conductance calculated from the numerical solution 
of Eq.~(\ref{psi-1}) is exactly identical to that computed in another 
numerical methods such as the recursive Green function method.~\cite{nonoyama,lee} 
In this paper, we solve Eq.~(\ref{psi-1}) within the single-site approximation, where
the multiple scattering effect involving many impurities (Anderson localization) are neglected. 
However the multiple scattering by an impurity is taken into account up to the 
infinite order of the scattering events. 
In the summation of $j$ in Eq.~(\ref{psi-1}), 
only the contribution with $j=j'$ is taken into account in the single-site 
approximation.~\cite{asano5}
In this way, the wave function at $\boldsymbol{r}_{j}$ is approximately
given by 
\begin{equation}
{\psi}^{(l)}(\boldsymbol{r}_{j}) \approx \left[ \hat{\sigma}_0
- \hat{G}_0^{SS}(\boldsymbol{r}_j,\boldsymbol{r}_j)\;V_i \;\hat{\sigma}_3\right]^{-1}
 {\psi}^{(l)}_0(\boldsymbol{r}_j).
\end{equation}
We note that the single-site approximation yields the exact conductance when $N_i=1$.
Within the single-site approximation, Eq.~(\ref{ls1}) can be solved as
\begin{align}
{\psi}^{(l)}_{\textrm{SSA}}(\boldsymbol{r}) 
=& {\psi}^{(l)}_0(\boldsymbol{r})
+ \sum_{j}^{N_i} \hat{G}_0^{NS}(\boldsymbol{r},\boldsymbol{r}_j)\;V_i \;\hat{\sigma}_3 
\nonumber\\ 
&\times \left[ \hat{\sigma}_0
- \hat{G}_0^{SS}(\boldsymbol{r}_j,\boldsymbol{r}_j)\;V_i \;\hat{\sigma}_3\right]^{-1}
 {\psi}^{(l)}_0(\boldsymbol{r}_j),\label{wf1}
\end{align}
for $x<0$. On the right hand side of Eq.~(\ref{wf1}), all functions have been 
given by analytical expressions.

In the presence of the impurity scattering,
the wave function Eq.~(\ref{wf1}) 
 can be expressed as
\begin{align}&{\psi}^{(l)}_{\textrm{SSA}}(\boldsymbol{r})=\left( \begin{array}{c} 1 \\ 0 \end{array}\right) \chi_l(y)e^{iq_l^+x} \nonumber\\
+&\sum_{l'} \chi_{l'}(y) \left[ \left( \begin{array}{c} 0 \\ 1 \end{array}\right) 
e^{iq^-_{l'}x}
A_{l',l} 
+\left( \begin{array}{c} 1 \\ 0 \end{array}\right) e^{-iq^+_{l'}x} B_{l',l}
\right],
\end{align}
for $x<0$, where $A_{l',l}$ and $B_{l',l}$ are the Andreev and 
the normal reflection coefficients in the presence of impurities, respectively.
These coefficients are obtained from relations
\begin{align}
\int_{-W/2}^{W/2} &\!\!\!\!\!\!
 dy\; \chi_m^\ast(y) (0, 1) {\psi}^{(l)}_{\textrm{SSA}}(\boldsymbol{r})
= e^{iq_m^- x} A_{m,l},\label{aml}\\
\int_{-W/2}^{W/2} &\!\!\!\!\!\! 
dy\; \chi_m^\ast(y) (1, 0) {\psi}^{(l)}_{\textrm{SSA}}(\boldsymbol{r})
= e^{iq^+_l x} \delta_{l,m} + e^{-iq^+_m x} B_{m,l}.\label{bml}
\end{align}
The scattering theory based on the Lippmann-Schwinger equation requires 
complicated algebra as shown below because the perturbation expansion
is carried out in the real space. 
In return, effects of the impurity scattering can be taken into account up to 
the infinite order of the perturbation expansion without using any 
self-consistent treatments. In addition, the reflection coefficients are
explicit functions of the impurity positions in a single disordered sample.
These are advantages of the present method.

\begin{figure}[htbp]
\begin{center}
\includegraphics[width=9.0cm]{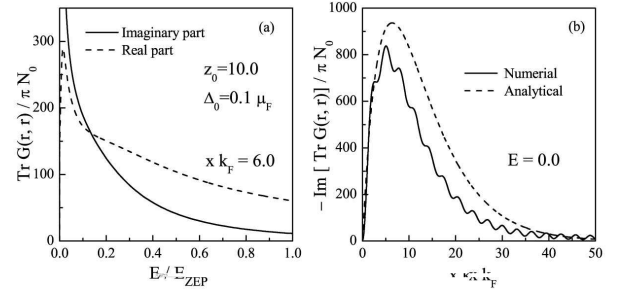}
\end{center}
\caption{
In (a), the trace of the Green function in the superconductor is shown as a function of $E$,
where $z_0=10$, $x k_F=6$ and $\Delta_0=0.1\mu_F$. 
The peak width of the imaginary part is given by
$E_{\textrm{ZEP}}=\Delta_0/z_0^2$. 
The energy scale, $E_{\textrm{dip}}$, characterizes the drastic increase 
of the imaginary part and the drastic decrease of the real part.
In (b), the local density of states is shown as a function of $x k_F$,
where $E=0$, $z_0=10$ and $\Delta_0=0.1\mu_F$. 
The numerical and the analytical results are denoted by the solid and the broken lines,
respectively.}
\label{fig:dos}
\end{figure}

In what follows, we consider low transparent junctions, (i.e., $z_0^2 \gg1$).
From the reflection coefficients in Appendix~A, the 
Green function in the superconductor is given by
\begin{widetext}
\begin{align}
\hat{G}_{0}^{SS}&(\boldsymbol{r},\boldsymbol{r}) = -i\pi N_0 
\frac{2}{\pi} \int_0^{\pi/2} \!\! d\gamma\; \left[ 
\frac{E}{2\Omega}\hat{\sigma}_0
-\frac{z_0^2 \Delta^2_{\boldsymbol{k}} \, e^{2ipx} }{2\Xi \Omega}\hat{\sigma}_0
+\frac{z_0^2 \Delta^2_{\boldsymbol{k}} 
\cos(2k_Fx \cos\gamma)\, e^{2ipx} }{2\Xi \Omega}\hat{\sigma}_0 \right. 
\nonumber \\
&-\frac{z^2_0 E}{2\Xi} \left\{ \frac{E}{\Omega} \cos(2k_Fx \cos\gamma)\hat{\sigma}_0 +i
\sin(2k_Fx \cos\gamma)\hat{\sigma}_3\right\}e^{2ipx}
- \frac{\Delta^2_{\boldsymbol{k}} \cos^2\gamma \, e^{2ipx}}{4\Omega\Xi}\hat{\sigma}_0
\nonumber\\
&- \left.
\frac{z_0 \cos\gamma}{2\Xi}e^{2ipx} 
\left\{E \cos(2k_Fx \cos\gamma) \hat{\sigma}_3\ +i \Omega \sin(2k_Fx \cos\gamma)\hat{\sigma}_0\right\}
\right],\label{g0jj}
\end{align}
\end{widetext}
where $p \approx \frac{k_F}{2\cos\gamma}\frac{\Omega}{\mu_F}.$
The local density of states~\cite{tanuma98} at $\boldsymbol{r}$ is defined by
\begin{equation}
N_s(E,x)= - \frac{1}{\pi} \textrm{Im} \textrm{Tr} 
\hat{G}_{0}^{SS}(\boldsymbol{r},\boldsymbol{r}). \label{dos-def}
\end{equation}
The first term in Eq.~(\ref{g0jj}) contributes to the bulk density of states.
Since $2p$ is roughly estimated to be $i/\xi_0$, another terms contribute to the 
local density of states near the interface, where $\xi_0=\hbar v_F/\pi \Delta_0$
is the coherence length and $v_F=\hbar k_F/m$ is the Fermi velocity.
In low transparent junctions, 6th, 7th and 8th terms are negligible.
The 4th and the 5th terms are also negligible 
because integrals of such rapidly oscillating functions become very small.
The 2nd and 3rd terms are dominant for $E\ll \Delta_0$.
In Fig.~\ref{fig:dos} (a), we show the trace of the Green function in Eq.~(\ref{g0jj}) 
as a function of $E$, where
$z_0$=10, $\Delta_0=0.1 \mu_F$, $x k_F=6$, and 
$N_0 = m/(\pi\hbar^2)$ is the normal density of states in the unit area.
Since $\xi_0k_F \sim 6.3$, the results correspond to
the Green function at a distance $\xi_0$ away from the interface.
The horizontal axis in Fig.~\ref{fig:dos} (a) is normalized by 
\begin{equation}
E_{\textrm{ZEP}}\equiv \frac{\Delta_0}{z_0^2}.
\end{equation}
The solid and the broken lines represent negative of the imaginary part and 
the real part of Eq.~(\ref{g0jj}), respectively.
As shown in Fig.~\ref{fig:dos} (a), the imaginary part of the Green function has a large peak 
around $E=0$ reflecting the ZES formed at the junction interface.
The energy scale, $E_{\textrm{ZEP}}$, characterizes the width of the 
zero-energy peak.
The real part of the Green function first increases with decreasing $E$ then
suddenly decreases to zero for $E\to 0$. The detail analysis indicates that 
the real part of the Green function has its maximum around an energy
\begin{equation}
E_{\textrm{dip}}\equiv \frac{\Delta_0}{z_0^2 (x k_F)^3}. \label{edip}
\end{equation}
The Green function for $E < E_{\textrm{ZEP}}$ is approximately 
calculated from the 2nd and the 3rd terms of Eq.~(\ref{g0jj})
\begin{align}
\hat{G}^{SS}_0&(\boldsymbol{r},\boldsymbol{r}) \approx  
2\pi N_0 z_0^2 e^{-x/\xi_0}( g_2 - i g_1)\hat{\sigma}_0,\\
g_1 =& \frac{2}{\pi} \int_0^{\pi/2}\!\!\!\! d\gamma \;
\frac{\Delta_0^2 \cos^4\!\gamma\, \sin^2\!\gamma \,\sin^2( x k_F\cos\gamma)}
{E^2z_0^4 + \Delta_0^2 \cos^6\gamma \sin^2\gamma},\\
g_2 =& \frac{2}{\pi} \int_0^{\pi/2}\!\!\!\! d\gamma \;
\frac{Ez_0^2 \Delta_0 \cos\gamma \, \sin\gamma \, \sin^2( x k_F\cos\gamma)}
{E^2z_0^4 + \Delta_0^2 \cos^6\gamma \sin^2\gamma},
\end{align}
where we use $\Omega \sim i |2\Delta_0 \cos\gamma \sin\gamma|$.
The imaginary part of the Green function, $g_1$, 
is of the order of unity when $E \sim E_{\textrm{ZEP}}$.
However, $g_1$ at $E=0$ becomes much larger than unity for $x k_F \gg 1 $ because
\begin{equation}
g_1 (E=0) = \frac{2}{\pi} \int_0^{\pi/2}\!\! d\gamma \;
\frac{ \sin^2\left\{x k_F\cos\gamma \right\}}
{ \cos^2\gamma} \sim x k_F. \label{e0dos}
\end{equation}
Thus the energy scale, $E_{\textrm{dip}}$, characterizes the drastic increase 
of $g_1$ and the drastic decrease of $g_2$.
The local density of states at $E=0$ calculated from Eqs.~(\ref{g0jj}) and (\ref{dos-def})
is plotted as a function of $x k_F$
in Fig.~\ref{fig:dos}(b) with the solid line. For comparison, we also show the analytical 
results represented by
\begin{equation}
-\left.\textrm{Im}\hat{G}_0^{SS}(\boldsymbol{r},\boldsymbol{r})\right|_{E=0} \simeq 2\pi N_0z_0^2 e^{-x/\xi_0} x k_F\hat{\sigma}_0, \label{g0dos2}
\end{equation}
with the broken line.
The results show the remarkable enhancement of the local density of states around $x\sim \xi_0$.
This implies that the ZES is formed around $x \sim \xi_0$.

Here we note several remarks as follows.
To calculate the Green function, we consider the 3rd term in Eq.(\ref{g0jj}) which rapidly
oscillates as $\cos(2xk_F\cos\gamma)$. Such rapidly oscillating terms
are usually neglected in the quasiclassical Green function method.
We, however, cannot neglect the 3rd term because it removes the divergence of the local density 
of state at $E=0$.~\cite{tanaka6,matsumoto} 
The 3rd term also becomes important when we
calculate the local density of states just at the surface, (i.e., $x=0$),
\begin{align} 
\frac{N_s(E,0)}{N_0} =& \textrm{Re} \frac{2}{\pi}
\int_0^{\pi/2}\!\!\!\!\!\!\!\! d\gamma \left[ \frac{E}{\Omega} - 
\frac{2E^2z_0^2 + \Delta^2_{\boldsymbol{k}}\cos^2\gamma}{2\Xi \Omega}\right], \\
\simeq  \frac{2}{\pi}K\!\!\left(\frac{\Delta_0}{E}\right)\!\!
&+ \frac{2}{\pi}
\int_0^{\pi/2}\!\!\!\!\!\!\!\! d\gamma
\frac{E^2z_0^2 + \Delta_0^2 \cos^6\gamma \sin^2\gamma}
{E^2z_0^4 + \Delta_0^2 \cos^6\gamma \sin^2\gamma},\label{dos00}
\end{align}
where $K(x)$ is the complete elliptic integral of the first kind and describes
the bulk density of states. Another terms come from the 4th and 6th terms in 
Eq.~(\ref{g0jj}).
The first equation is the exact expression and 
 we use $E < E_{\textrm{ZEP}}$ in the second line.
We exactly obtain $N(E=0,x=0)=N_0$. Thus there is no remarkable enhancement in the
zero energy local density of states just at the interface.
The 2nd and the 3rd terms in Eq.~(\ref{g0jj}) do not contribute
to the $N_s(E,0)$ because they exactly cancel with each other at $x=0$.

In the next section, the conductance for $E<E_{\textrm{ZEP}}$ will be discussed.
It is possible to rewrite a part of Eq.~(\ref{wf1}) as
\begin{align}
&\left[ 
\hat{\sigma}_0 
- V_i \hat{G}^{SS}_0(\boldsymbol{r},\boldsymbol{r}) 
\hat{\sigma}_3 
\right]^{-1} 
= \frac{\hat{\sigma}_0 + s \hat{\sigma}_3}{1-s^2},\label{self}\\
s_1 &=  Q_0 \; g_1,\label{s1}\\
s_2 &=  Q_0 \; g_2,\label{s2}\\
Q_0&= 2\pi V_i N_0 z_0^2 e^{-x/\xi_0}\label{q0},\\
s &= s_2-i s_1,
\end{align}
where $s$ corresponds to the self-energy of a quasiparticle scattered by an impurity once 
and describes two important features.
First, $s$ becomes large in low transparent
junctions even if $V_iN_0$ is fixed at a small constant because 
$Q_0$ in Eq.(\ref{q0}) is proportional to $V_iN_0z_0^2 \sim V_iN_0/T_B$.
Thus $Q_0$ represents the normalized strength of the impurity scattering.
This behavior explains the previous numerical simulation.~\cite{asano2}
Second, effects of impurity scattering far away from the interface on the conductance 
is negligible because $Q_0$ decreases exponentially with the increase of $x$.
Impurities around the ZES, (i.e.,  $x \sim \xi_0$) seriously affect the conductance 
for $E<E_{\textrm{ZEP}}$.

\section{conductance}
The differential conductance in NS junctions is calculated from the 
normal and the Andreev reflection coefficients,~\cite{blonder,takane}
\begin{align}
G_{NS}&(eV) = 
\frac{2e^2}{h}\sum_{l,m} \int_{-\infty}^\infty\!\!dE \; 
\left( \frac{\partial f_{\textrm{FD}}(E-eV)}{\partial (eV)} \right)\nonumber\\
&\times\left[ \delta_{l,m} - |\bar{B}_{m,l}(E)|^2 + |\bar{A}_{m,l}(E)|^2 \right],\\
\bar{A}_{m,l} =& \sqrt{\frac{k_m}{k_l}}A_{m,l},\\
\bar{B}_{m,l} =& \sqrt{\frac{k_m}{k_l}}B_{m,l},
\end{align}
where $f_{\textrm{FD}}(E)$ is the Fermi-Dirac distribution function.
When $z_0^2 \gg 1$ and $E < E_{\textrm{ZEP}}$, the reflection coefficients are calculated as
\begin{align}
\bar{A}_{m,l} =& \delta_{l,m} r_{NN}^{he}(l)\nonumber\\
 + &e^{-i\varphi}L_{m,l}
\left[ \Delta_m |\Delta_l| (s+1) + \Delta_l |\Delta_m|(s-1)\right], \\
\bar{B}_{m,l} =& \delta_{l,m} r_{NN}^{ee}(l) \nonumber\\
&+ i L_{m,l}
\left[ |\Delta_m| |\Delta_l| (s+1) + \Delta_l \Delta_m(s-1)\right],\\
L_{m,l}=& \frac{\pi N_0 V_i z_0^2 }
{(1-s^2)k_F}
 \sum_{j=1}^{N_i} \chi_l(y_j) \chi^\ast_m(y_j) \nonumber\\ 
&\times\frac{\sqrt{\bar{k}_l \bar{k}_m}}{\Xi_l \Xi_m}
e^{i (p_m +p_l)x_j}\sin(k_mx_j) \sin(k_l x_j). 
\end{align}
The conductance is then given by
\begin{align}
&G_{NS} = \frac{2e^2}{h}\int_{-\infty}^\infty\!\!dE \; 
\left( \frac{\partial f_{\textrm{FD}}(E-eV)}{\partial (eV)} \right)\nonumber\\
&\times \left[ N_c g^{(0)} 
 -  4 \sum_{j=1}^{N_i} \Gamma_j \right],\label{gns2}\\
&g^{(0)}= 2\int_0^{\pi/2}\!\!\!\!d\gamma\; 
\frac{ \Delta_0^2\cos^7\!\gamma \sin^2\!\gamma}{E^2z_0^4 + 
\Delta_0^2\cos^6\!\gamma \sin^2\!\gamma},\label{gclean}\\
\Gamma_j =& \textrm{Re} \frac{ s Q_0 (2I_2-iI_1+iI_3)}{s^2-1},\label{gam1}\\
I_1=& \frac{2}{\pi} \int_0^{\pi/2} 
\!\!\!\!\! d\gamma\; 
\frac{ \Delta_0^4\cos^{10}\!\gamma \sin^4\!\gamma \sin^2\!( x_jk_F\cos\gamma)}
{(E^2z_0^4 + \Delta_0^2\cos^6\!\gamma \sin^2\!\gamma)^2},\\
I_2=& \frac{2}{\pi} \int_0^{\pi/2} 
\!\!\!\!\! d\gamma\; 
\frac{ Ez_0^2 \Delta_0^3\cos^{7}\!\gamma \sin^3\!\gamma \sin^2\!( x_jk_F\cos\gamma)}
{(E^2z_0^4 + \Delta_0^2\cos^6\!\gamma \sin^2\!\gamma)^2},\\
I_3=& \frac{2}{\pi} \int_0^{\pi/2} 
\!\!\!\!\!\!\! d\gamma 
\frac{ E^2 z_0^4 \Delta_0^2\cos^{4}\!\gamma \sin^2\!\gamma \sin^2\!( x_jk_F\cos\gamma)}
{(E^2z_0^4 + \Delta_0^2\cos^6\!\gamma \sin^2\!\gamma)^2}.
\end{align}
The first term of Eq.~(\ref{gns2}), $N_cg^{(0)}$, is the conductance in clean
junctions and
$\Gamma_j$ represents effects of the impurity scattering on the conductance.
When we calculate $|\bar{A}_{m,l}|^2$ and $|\bar{B}_{m,l}|^2$, 
the summation with respect to impurities $\sum_j^{N_i}\sum_{j'}^{N_i}$
must be carried out only for $j'=j$ in the single-site approximation.~\cite{asano5} 
As a consequence, the current conservation law is satisfied for $E<E_{\textrm{ZEP}}$.

To study effects of impurities on the conductance, we first assume $x_j=x_0$
for all impurities. The conductance is rewritten as
\begin{align}
G_{NS} =& \frac{2e^2}{h}\int_{-\infty}^\infty\!\!dE \; 
\left( \frac{\partial f_{\textrm{FD}}(E-eV)}{\partial (eV)} \right)\nonumber\\
&\times N_c \left[ g^{(0)}  - 2 n_i \Gamma_j \right],\label{gns3}
\end{align}
where $n_i = N_i \lambda_F/W$ is the dimensionless line density
 of impurities less than unity. 
When scattering effects are strong, $|s| \gg 1$,
$N_i$ cannot be much larger than $W/\lambda_F$. This limits the applicability
of the single-site approximation. 

\begin{figure}[htbp]
\begin{center}
\includegraphics[width=9.0cm]{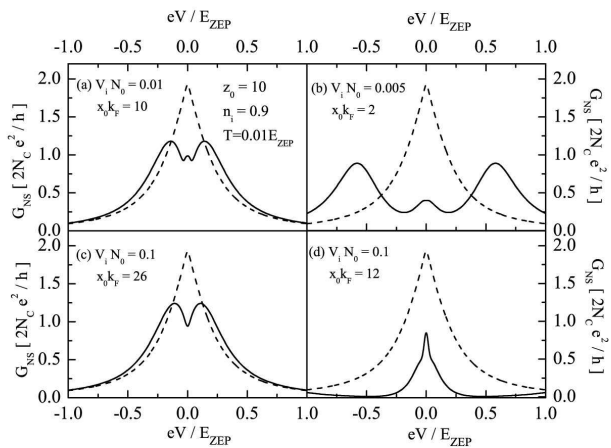}
\end{center}
\caption{
The conductance is plotted as a function of bias voltages,
where $z_0=10$ and $n_i=0.9$. The parameters are 
$x_0k_F=10$, $V_i N_0 =0.01$ in (a), $x_0k_F=2.0$,
 $V_i N_0 =0.005$ in (b), $x_0k_F=26$, $V_i N_0 =0.1$ in (c) and
 $x_0k_F=12$, $V_i N_0 =0.1$ in (d).
The broken lines denote the conductance in clean junctions.  
}
\label{fig:gline}
\end{figure}

\begin{figure}[htbp]
\begin{center}
\includegraphics[width=8.0cm]{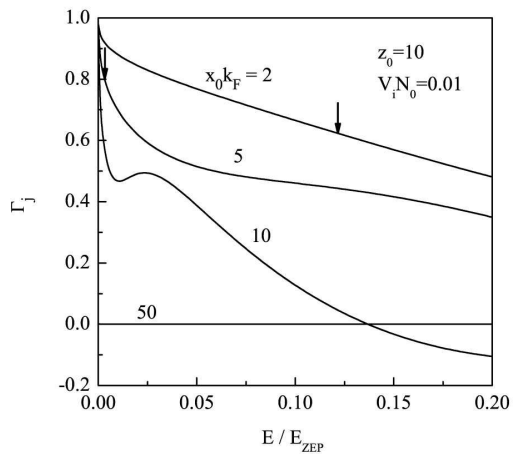}
\end{center}
\caption{
The function $\Gamma_j(E)$ is plotted as a function of $E$ for several $x_0k_F$ at 
$V_iN_0=0.01$. The arrows indicate $E_{\textrm{dip}}$. For $x_0k_F=10$, $E_{\textrm{dip}}$
= 0.001 $E_{\textrm{ZEP}}$. 
}
\label{fig:gamma}
\end{figure}

We show conductance for several choices of $x_0k_F$ and $V_iN_0$ in Fig.~\ref{fig:gline}, where 
$z_0=10$ and $n_i=0.9$. The two parameters are chosen as 
$x_0k_F=10$, $V_i N_0 =0.01$ in (a), $x_0k_F=2.0$,
 $V_i N_0 =0.005$ in (b),  $x_0k_F=26.0$, $V_i N_0 =0.1$ in (c) 
and $x_0k_F=12.0$, $V_i N_0 =0.1$ in (d).
The broken line is the conductance in clean junctions.
The temperature is fixed at a very low temperature $T=0.01E_{\textrm{ZEP}}$
which is estimated to be 0.05 K by using $\Delta_0=50$ meV for $z_0=10$. 
 As shown in (a)-(c), the ZBCP is splitting into two peaks by the impurity scattering.
While the results in (d) shows the single ZBCP.
Roughly speaking,
the impurity scattering affects the ZBCP in two ways:
(i) it decreases the conductance around the zero bias voltage and (ii) it makes the ZBCP wider.
The two effects (i) and (ii) are well characterized by the $E$ dependence of $\Gamma_j$ and
the sign change of $\Gamma_j$ in Eq.~(\ref{gns3}), respectively.
In Fig.~\ref{fig:gamma}, $\Gamma_j$ is plotted as a function of $E$ for several $x_0k_F$,
where $V_iN_0$ is fixed at 0.01. We note that $\Gamma_j$ for $x_0k_F=10$ yields the
conductance in Fig.~\ref{fig:gline} (a).
When $x_0k_F \gg \xi_0$, $\Gamma_j \sim 0$ and impurity scattering is negligible as shown 
for $x_0k_F=50$ because $Q_0$ in Eq.~(\ref{q0}) becomes almost zero. 
For $x_0k_F=$ 2, 5 and 10, $\Gamma_j$ increases with decreasing $E$,
which indicates the enhancement of the impurity scattering around $E=0$.
The suppression of the conductance around the zero bias is explained in terms of 
the drastic increase of the local density of states with decreasing $E$
as shown in Fig.~\ref{fig:dos} (a). Therefore the suppression of the zero-bias conductance 
happens irrespective of $x_0k_F$ and $V_iN_0$.
In addition, a nonmonotonic $E$ dependence of $\Gamma_j$ for $x_0k_F=10$ is a source
of small conductance peak at the zero-bias in Fig.~\ref{fig:gline} (a). The same
small peak is also found in Fig.~\ref{fig:gline} (b). 

The widening of the ZBCP can be explained by the sign change of $\Gamma_j$. 
When $E>0.15E_{\textrm{ZEP}}$ in Fig.~\ref{fig:gamma}, $\Gamma_j$
for $x_0k_F=10$ becomes negative and impurities enhance the conductance. 
As a consequence, the conductance peak becomes wider than $g^{(0)}$.
The split of the ZBCP is a consequence of the interplay between the suppression of 
the conductance around the zero bias voltage and the widening of the ZBCP 
as shown in Fig.~\ref{fig:gline} (a)-(c).
Therefore the sign change of $\Gamma_j$ explains the split of the ZBCP.  
For $E_{\textrm{dip}} < E < E_{\textrm{ZEP}}$,
$\Gamma_j$ is almost a decreasing function of $E$ and is positive
at $E=E_{\textrm{dip}}$ as shown in Fig.~\ref{fig:gamma}. 
It is followed from Eq.~(\ref{gam1}) that 
\begin{equation}
\Gamma_j \propto |s|^2\left\{ s_1(I_1-I_3) + 2I_2s_2 \right\}+
\left\{ s_1(I_1-I_3) - 2I_2s_2 \right\}.
\end{equation}
Within our study, $s_1(I_1-I_3)$ tends to be much smaller than $2I_2s_2$ for 
$E_{\textrm{dip}} < E < E_{\textrm{ZEP}}$, which implies an importance
of the real part of the self-energy, $s_2$, for the splitting. 
The sign change of $\Gamma_j$ happens when the impurity scattering is sufficiently
weak so that 
\begin{equation}
|s|^2=s_1^2+s_2^2 \sim 1, \label{sp-cond}
\end{equation}
is satisfied. 
When $|s(E=E_{\textrm{dip}})|$ is a small value less than unity, 
the effects of impurities are negligible and the conductance almost remains unchanged 
from that in clean junctions.
The split also cannot be seen when $|s(E=E_{\textrm{ZEP}})|$ is larger than unity.
An example is shown in Fig.~\ref{fig:gline}(d), where $x_0k_F=12.0$, 
$V_i N_0 =0.1$ and $|s(E=E_{\textrm{ZEP}})|$ is estimated to be 1.5. 
In this case, the suppression of the zero-bias conductance dominates over
the widening of the ZBCP. As a result,
the conductance is always smaller than that in clean junctions and the ZBCP remains in
a single peak.

 We should pay attention to a similarity in the shapes of the conductance 
in the present theory and those in experiments.
Amazingly, the conductance structure in Fig.~\ref{fig:gline} (a) is very similar to that 
observed in the experiment~\cite{covington}. It is possible to find a very small conductance
peak at $V=0$ in addition to the splitting peaks around $V \sim \pm 1$mV in Fig.~2
of Ref.~\onlinecite{covington}. 
The impurities away from the interface explains another conductance shape in the
experiment~\cite{iguchi}.
The conductance structure in Fig.~2 of Ref.~\onlinecite{iguchi} is very similar
to that in Fig.~\ref{fig:gline} (d).
The present theory explains, at least, two 
typical conductance shapes observed in the experiments.

\begin{figure}[htbp]
\begin{center}
\includegraphics[width=8.5cm]{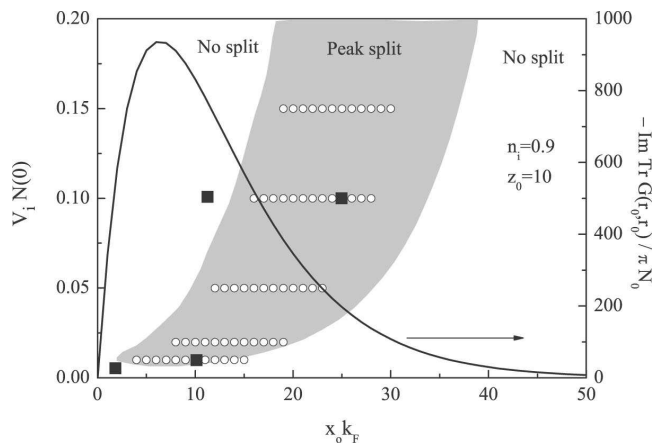}
\end{center}
\caption{
A phase diagram for the split of the ZBCP.
The gray area indicates sets of ($V_iN_0, x_0k_F$) which satisfy
Eq.~(\ref{sp-cond}). The open circles denote sets of ($V_iN_0, x_0k_F$), where we 
find the split of the ZBCP in Eq.~(\ref{gns3}). The local density of state at $E=0$
in Fig.~\ref{fig:dos} (b) is also shown. The filled squres are parameters used for the 
conductance in Fig.~\ref{fig:gline}.
}
\label{fig:diagram}
\end{figure}

As shown in Fig.~\ref{fig:gline}, the magnitude of the impurity potential and 
the position of impurities are key factors for the degree of splitting. 
In Fig.\ref{fig:diagram}, the gray area indicates 
sets of ($V_iN_0, x_0k_F$) which satisfy
Eq.~(\ref{sp-cond}) within $E_{\textrm{dip}} < E < E_{\textrm{ZEP}}$.
The open circles denote sets of ($V_iN_0, x_0k_F$), where we find
the split of the ZBCP in Eq.~(\ref{gns3}).
All the circles are inside of the gray area. 
Although the circles and the gray region are not perfectly coincide 
with each other, they show qualitatively the same tendency. 
The parameters used in Fig.~\ref{fig:gline} are indicated by filled squares.
Since $s$ at $E=0$ is proportional to the local density of states, 
we also show the $N_s(E=0,x)$ in Eq.~(\ref{g0dos2}).
To satisfy Eq.~(\ref{sp-cond}), impurities around $x_0\sim\xi_0$ 
should have sufficiently small scattering potentials because the local density
of states has large values there.
Thus the gray area appears for small impurity potentials near the interface, 
(i.e., $x_0 \lesssim \xi_0$).
 The gray region spreads to larger $V_iN_0$ as the increase of $x_0 > \xi_0$
because the local density of states becomes smaller values. 
The results imply that the strong impurities are not necessary for the split of
the ZBCP. 
It is evident that this phase diagram is valid in the limit of high impurity density
and the diagram would be changed depending on the transmission probability of junctions. 

\begin{figure}[htbp]
\begin{center}
\includegraphics[width=8.5cm]{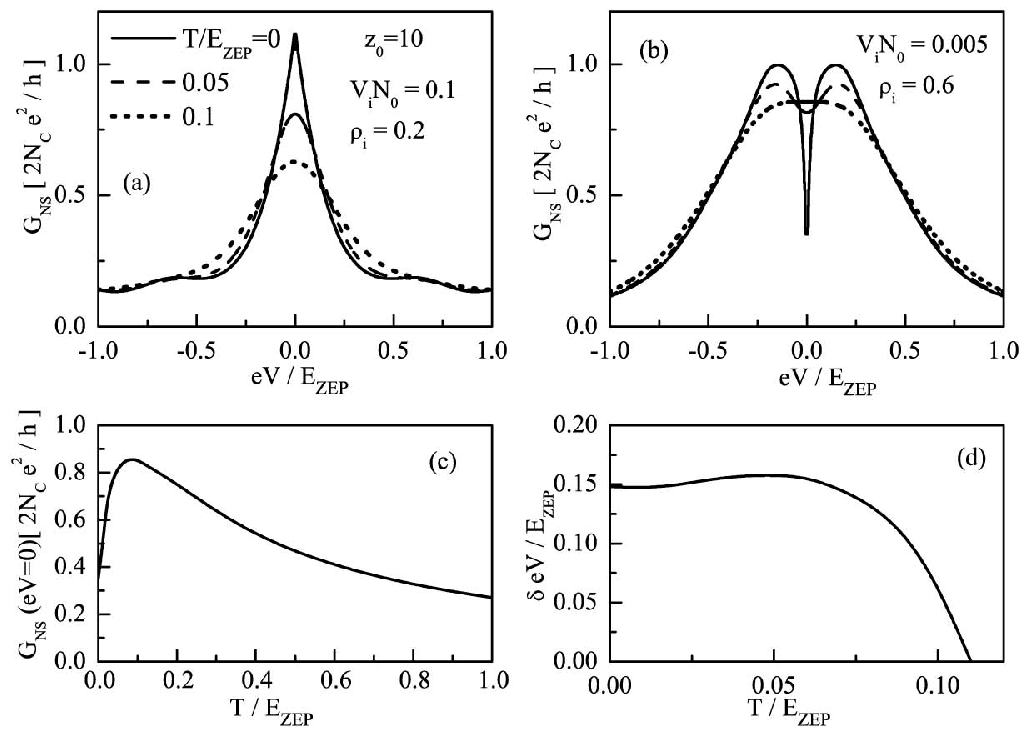}
\end{center}
\caption{
The conductance in the presence of impurities distributed randomly 
in the range of $1 <x_jk_F < 20$, where  
$\rho_i$ is the dimensionless area density of impurities near the interface.  
The conductance for low density strong impurities is shown in (a) with
$V_iN_0=0.1$ and $\rho_i=0.2$. The conductance for
high density weak impurities with $V_iN_0=0.005$ and $\rho_i=0.6$ are
shown for several choices of temperatures.
The zero-bias conductance and the peak positions in (b) 
are plotted as a function of temperatures in (c) and (d), respectively.
}
\label{fig:tdep}
\end{figure}

In Fig.~\ref{fig:gline}, all impurities are aligned at $x_j=x_0$.
In real junctions, however, impurities may be distributed randomly near the interface
as shown in Fig.~\ref{system}. 
The conductance in such realistic junctions are shown in Fig.~\ref{fig:tdep}, where
impurities are distributed randomly in the range of $1 <x_jk_F < L_s k_F$,
$\rho_i=N_i\frac{\lambda_F^2}{WL_s}$ is the dimensionless area density of impurities
and $z_0=10$. 
The conductance is calculated from an expression
\begin{align}
G_{NS} =& \frac{2e^2}{h}N_c \int_{-\infty}^\infty\!\!dE \; 
\left( \frac{\partial f_{\textrm{FD}}(E-eV)}{\partial (eV)} \right)\nonumber\\
& \times \left[ g^{(0)} 
  -  \rho_i \frac{k_FL_s}{\pi} \left\langle \Gamma \right\rangle \right],\label{gns4}\\
\left\langle \Gamma \right\rangle=& \left\langle\frac{1}{N_i} \sum_{j=1}^{N_i} \Gamma_j
\right\rangle,
\end{align}
where $\langle \cdots \rangle$ represents the ensemble average.
Since the conductance in Eq.~(\ref{gns2}) is characterized by 
the number of impurities, a factor $k_FL_s/\pi$ appears in Eq.~(\ref{gns4}).
We choose $L_sk_F=20$ in Fig.~\ref{fig:tdep} because we focus on the 
impurities near the interface and $\xi_0k_F \sim 6.3$.
We consider low density strong impurities in (a), where $V_iN_0=0.1$ 
and $\rho_i=0.2$.
There is no peak splitting in Fig.~\ref{fig:tdep} (a) because most impurities are outside 
of the gray region in Fig.~\ref{fig:diagram}.
On the other hand, the results in Fig.~\ref{fig:tdep} (b) show the 
split of the ZBCP at $T=0$, where 
we consider high density weak impurities with for $V_iN_0=0.02$ and $\rho_i=0.6$.
This is because most impurities are inside of the gray region in Fig.~\ref{fig:diagram}.
The splitting peaks merge into a single peak under finite temperatures such
as $T=0.1E_{\textrm{ZEP}}$. In Fig.~\ref{fig:tdep} (c), we show the 
temperature dependence of the zero-bias
conductance in Fig.~\ref{fig:tdep} (b). The results show the reentrant behavior of the zero-bias 
conductance, which was found in the experiment.~\cite{covington} 
In Fig.~\ref{fig:tdep} (d), the peak position ($\delta eV$) in (b) 
is plotted as a function temperatures.
Since $\delta eV$ is about 0.15 $E_{\textrm{ZEP}}$ at $T=0$, peak splitting is washed out
at high temperatures such as $T=0.11 E_{\textrm{ZEP}}$.
High density impurities with weak random potential are responsible
for the split of the ZBCP in low transparent junctions.

\begin{figure}[htbp]
\begin{center}
\includegraphics[width=8.5cm]{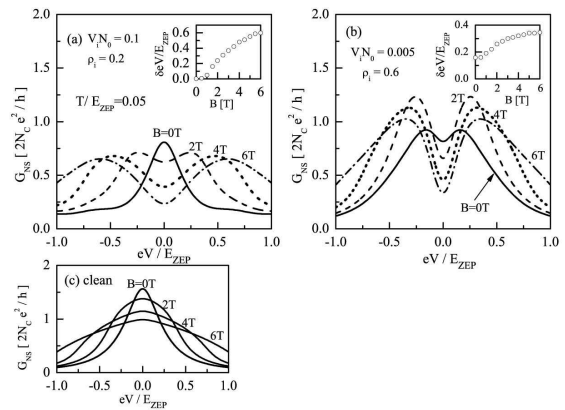}
\end{center}
\caption{
The conductance under external magnetic fields for low density strong impurities
with $V_iN_0=0.1$ and $\rho_i=0.2$ are shown in (a), where $T$ is fixed at $0.05 E_{\textrm{ZEP}}$.
Those for high density weak impurities with $V_iN_0=0.005$ and $\rho_i=0.6$ are
shown in (b).
In insets, peak positions are plotted as a function of magnetic fields.
The conductance of clean junctions is shown in (c).
}
\label{fig:bdep}
\end{figure}

Several experiments~\cite{covington,greene} show a sensitivity of the conductance peaks to external
magnetic fields. Here we discuss the conductance in the presence of magnetic fields.
The effects of magnetic fields are taken into account phenomenologically 
by using the Aharonov-Bohm like phase shift~\cite{ab1,zes} of a quasiparticle.
Since the impurity scattering in magnetic fields itself is a difficult problem 
to solve analytically, 
we neglect the interplay between magnetic fields and impurity scatterings.
Within the phenomenological theory,~\cite{zes} 
effects of magnetic fields is considered by replacing $E$ in Eq.~(\ref{gclean})
by $E+ |\Delta_0 \cos\gamma\sin\gamma|\phi_B$ as 
\begin{align}
g^{(0)}_B=& \int_{-\pi/2}^{\pi/2}\!\!\!\!d\gamma\; 
\frac{ \Delta_0^2\cos^7\!\gamma \sin^2\!\gamma}{E_B^2z_0^4 + 
\Delta_0^2\cos^6\!\gamma \sin^2\!\gamma},\label{gcleanb}\\
E_B=&E+ 2\Delta_0 |\cos\gamma\sin\gamma| \phi_B, \\
\phi_B =& 2\pi \frac{B \xi_0^2}{\phi_0} \tan\gamma = B_0\tan\gamma,
\end{align}
where $\phi_0=2\pi \hbar c/e$ and $B_0=1.0^{-3}$ corresponds to $B =1$ Tesra.
A quasiparticle acquires the Aharonov-Bohm like phase shift $\phi_B$ while
moving near the NS interface.~\cite{zes}
In a previous paper, we found that ZBCP in clean junctions remains a single peak
even in the strong magnetic fields~\cite{zes} as shown in Fig.~\ref{fig:bdep} (c),
where $z_0=10$ and $T=0.05E_{\textrm{ZEP}}$. 
In Fig.~\ref{fig:bdep} (a) and (b), we show the conductance in the presence of
low density strong impurities and high density weak impurities, respectively,
where $V_i$ and $\rho_i$ are same as those in Fig.~\ref{fig:tdep} (a) and (b), respectively.
A temperature is fixed at $T=0.05E_{\textrm{ZEP}}$. In contrast to clean junctions in 
Fig.~\ref{fig:bdep} (c),
the ZBCP in disordered junctions splits into two peaks under magnetic fields as 
shown in Fig.~\ref{fig:bdep}(a). 
The results within the phenomenological theory indicate that the sensitivity 
of the ZBCP to magnetic fields depends on the degree of impurity scatterings. 
In insets, peak positions $(\delta eV)$ are plotted with circles as a function of magnetic
fields. For high density weak impurities in Fig.~\ref{fig:bdep} (b), we also 
found that the degree of 
peak splitting increases with increasing magnetic fields. In the limit of the strong fields,
$\delta eV$ tends to saturate as shown in the inset. These characteristic behavior are found in the 
experiment.~\cite{covington}
Although we have well explained the characteristic behavior of the experimental 
conductance peaks under 
magnetic fields, the applicability of the phenomenological theory in the presence of 
impurities is still unclear.
This issue would be addressed more clearly in an exact numerical simulation.

\section{discussion}

In experiments, the split of the ZBCP has been reported in overdoped high-$T_c$
superconductors.~\cite{sharoni} 
The heavy carrier doping may bring a number of defects or imperfections in superconductors. 
It is evident that the impurity scattering is unavoidable
 even in underdoped high-$T_c$ superconductors.
The split of the ZBCP would be found in underdoped superconductors 
if they have bad sample quality. 
In the same way, the split of the ZBCP would not be found even 
in overdoped superconductors if their sample quality are good enough.
The sample quality is a key factor for the spilt of the ZBCP.
This argument is consistent with an experiment,~\cite{aprili} where
the potential disorder is artificially introduced to the NS junctions 
by the ion-irradiation and washes out the ZBCP in the limit of strong disorder.

In order to compare the theoretical results in this paper with experiments,
we may consider the impurity scattering in normal metals.
The total resistance ($R$) in the dirty normal-metal / (110) $d$-wave 
junction can be described simply by $R = R_D + R_{NS}$,~\cite{circuit,h-kashiwaya} 
where $R_D = 1/G_D$ is the resistance 
of the dirty normal metal and $G_{NS}=1/R_{NS}$ is the conductance discussed in this 
paper. The equation indicates the absence of the proximity effect in the dirty 
normal metal.~\cite{asano7,asano6} The height of the ZBCP can be reduced to 
reasonable values in the presence of the impurity scattering in normal metals 
because the total conductance is given by $1/(R_D + R_{NS})$.
The degree of splitting depends on parameters such as the potential of impurities ($V_j$),
the position of impurities ($x_j$) and the transparency of the junction 
($T_B\sim 1/z_0^2$).
In particular, $T_B$ is the most important parameter.
As shown in Figs.~\ref{fig:gline} and \ref{fig:tdep}, 
the degree of splitting is roughly given by 
$E_{\textrm{ZEP}}=\Delta_0/z_0^2 \sim \Delta_0 T_B$. 
Thus it is possible to choose $T_B$ to fit the degree of splitting with that found in experiments. 
The amplitude of the pair potential is about 30-40 meV in typical high-$T_c$ materials.
The degree of splitting is then estimated to be 1.2-1.6 meV for $z_0 = 5$, which is 
almost consistent with that found in experiments, for instance, 2 meV.~\cite{covington} 
It is also necessary to consider electronic structures of high-$T_c$ materials 
for the quantitative agreement of the splitting width in theories with that in experiments. 

In quasiclassical Green function theories, the conductance is proportional to the density
of states at the surface of superconductors.
We find that $E$ dependence of the density of states at the interface in 
Eq.~(\ref{dos00}) is apparently different from 
that of the conductance even in the clean junctions shown in Eq.~(\ref{gclean}). 
We also find that $N_s(E,x=0)\sim N_0 \ll N_s(E,x=\xi_0)$ as shown in Fig.~\ref{fig:dos} (b).
The density of states near the interface averaged over $\xi_0$ in this paper may 
corresponds to the surface density of states in the quasiclassical approximation, where 
the rapidly oscillating part of the wave function are 
neglected and the smallest length scales is given by $\xi_0$. 
It is impossible to directly compare the present theory with 
the quasiclassical Green function theories because
the position of impurities is a key parameter for the split of the ZBCP in our 
theory. 

The Abrikosov-Gor'kov (AG) theory~\cite{agd} is a useful approach to discuss 
the impurity scattering in superconductors. The applicability of the AG theory
is limited to the superconductivity in diffusive metals, 
where the mean free path is much smaller than the size of the disordered region
in superconductors.
Here we briefly discuss a relation between the AG theory and ours.
Since the AG theory assumes the diffusive transport regime, we define the mean free
path of a quasiparticle in the Born approximation,
\begin{align}
\ell =& v_F \tau,\\
\frac{\hbar}{\tau}=& 2\pi \bar{\rho}_i N_0V_i^2, \\
\bar{\rho}_i =& N_i/WL_s,
\end{align}
where we consider $W\times L_s$ disordered region on two-dimensional superconductor
as in Fig.~\ref{fig:tdep}. The impurity density $\bar{\rho}_i$ can be replaced by
the dimensionless impurity density $\rho_i =  (2\pi/k_F)^2 \bar{\rho}_i$.
The ratio of the mean free path and the coherence length 
is given by
\begin{equation}
\frac{\ell}{\xi_0}= \frac{\Delta_0}{\mu_F} \frac{\pi}{\rho_i (N_0V_i)^2}
\end{equation}
In this paper, $\ell/\xi_0$ is larger than unity even in the limit of 
$\rho_i=1$ because we assume $\Delta_0/\mu_F = 0.1$ and $N_0V_i < 0.1$. 
On the other hand, the dirty limit is defined by a relation 
$\ell/\xi_0 \ll 1$. 
When we choose $L_s$ being a few coherence length as shown in Fig.~\ref{fig:tdep},
where $L_s \sim 3 \xi_0$, we find that $\ell / L_s$ is still larger than unity.
The disordered region is not in the diffusive regime 
but in the quasi-ballistic regime because 
the diffusive regime is characterized by a relation $\ell / L_s \ll 1$.
Thus it is basically impossible to apply the Abrikosov-Gor'kov theory to the model in this paper. 
The scattering theory~\cite{asano5} used in this paper is a suitable 
analytic method to discuss the conductance of such NS junctions.
Although the impurity scattering near the interface is very weak in normal states, 
it drastically affects the conductance below the critical temperature.
Impurities located at the resonant states seriously suppress
the degree of the resonance even if their potentials are weak. 
 In the AG theory, the real part of the self-energy is usually neglected. 
In our theory, this approximation corresponds to an equation $s_2=0$. 
However $s_2$ plays an important role
in the peak splitting. When we omit $s_2$, $I_2$ and $I_3$ must be also
zero, which leads to positive $\Gamma_j$ and no splitting irrespective of
$V_iN_0$ and $x_jk_F$. 

We show that the impurity scattering causes the splitting of the ZBCP.
The conclusion, however, does not deny a possibility of the BTRSS.
In a previous paper,~\cite{kitaura} we assume the $s+id$ symmetry near the NS interface and 
numerically study the tunneling conductance. The results show that the split of the ZBCP
is insensitive to the potential disorder. Thus peak splitting would be always expected 
in low transparent junctions~\cite{Tanuma2001} if the BTRSS appears at the NS interface.
At present, we have only limited information on the BTRSS within the mean-field theories.
To understand the nature of the BTRSS further beyond the mean-field theory, 
we have to analyze electronic structure 
of high-$T_c$ superconductors based on microscopic models 
 and make clear effects of the surface, the electron correlation 
and the random potentials on the superconducting state. 
This is an important future problem.

Since the formation of the ZES is a universal phenomenon in superconductors with 
unconventional pairing symmetries, the ZES is also expected at a surface of spin-triplet
superconductors.~\cite{Buch} It is interesting to study effects of impurities 
on transport properties in spin-triplet superconductor 
junctions.~\cite{Yama1,Yama2,Yama3,asano2003-3,Y1,H1,H2,Kuroki1,Kuroki2,Kusakabe,Honer,St,Sen,asano6,asano8,asano9} 

\section{conclusion}
We have discussed effects of impurity scatterings on the conductance
in normal-metal/$d$ wave superconductor junctions.
The conductance is calculated from the Andreev and the normal reflection
coefficients which are estimated by using the single-site approximation.
We consider impurities near the junction interface on the superconductor side.
The strength of the impurity scattering strongly 
depends on the transparency of the junction, the position of impurities and the energy
of a quasiparticle because the ZES are formed at the NS interface. 
We conclude that the impurity scattering causes the split of the zero-bias
conductance peak. 
The results are consistent with previous numerical simulations.
We have also shown that characteristic behaviors of the conductance spectra at finite 
temperatures and under external
magnetic fields qualitatively agree with those reported in experiments.

\appendix
\section{Transmission and reflection coefficients}
In the clean NS junctions, the transmission and the reflection coefficients 
can be calculated from the appropriate boundary condition of the wave function.
The calculated results are shown below. 
\begin{align}
r^{he}_{NN}(l) = & \frac{ \bar{k}_l^2}{\Xi_l}\; \frac{\Delta_l}{2}\;  e^{-i\varphi},\\
r^{ee}_{NN}(l) = & \frac{ -iz_0(\bar{k}_l-iz_0)}{\Xi_l}\; E,\\
t^{ee}_{SN}(l) = & \frac{\bar{k}_l(\bar{k}_l-iz_0)}{\Xi_l}\; E\; u_l \; e^{-i\varphi/2},\\
t^{he}_{SN}(l) = & \frac{iz_0 \bar{k}_l}{\Xi_l}\; E\; v_l \; e^{-i\varphi/2},
\end{align}
\begin{align}
r^{eh}_{NN}(l) = & \frac{-\bar{k}_l^2}{\Xi_l}\; \frac{\Delta_l}{2}\;  e^{i\varphi},\\
r^{hh}_{NN}(l) = & \frac{ iz_0(\bar{k}_l+iz_0)}{\Xi_l}\; E,\\
t^{hh}_{SN}(l) = & \frac{\bar{k}_l(\bar{k}_l+iz_0)}{\Xi_l}\; E\; u_l\;  e^{i\varphi/2},\\
t^{eh}_{SN}(l) = & \frac{iz_0 \bar{k}_l}{\Xi_l}\; E\; v_l \; e^{i\varphi/2},
\end{align}
\begin{align}
r^{he}_{SS}(l) = & \frac{\bar{k}_l^2+2z_0^2}{\Xi_l}\; \frac{\Delta_l}{2},\\
r^{ee}_{SS}(l) = & \frac{ -iz_0(\bar{k}_l-iz_0)}{\Xi_l}\; \Omega_l,\\
t^{ee}_{NS}(l) = & \frac{\bar{k}_l(\bar{k}_l-iz_0)}{\Xi_l}\; \Omega_l\; u_l \; 
e^{i\varphi/2},\\
t^{he}_{NS}(l) = & \frac{-iz_0 \bar{k}_l}{\Xi_l}\; \Omega_l \; v_l \; e^{-i\varphi/2},\\
\end{align}
\begin{align}
r^{eh}_{SS}(l) = & \frac{-(\bar{k}_l^2+2z_0^2)}{\Xi_l}\; \frac{\Delta_l}{2},\\
r^{hh}_{SS}(l) = & \frac{ iz_0(\bar{k}_l+iz_0)}{\Xi_l}\; \Omega_l,\\
t^{hh}_{NS}(l) = & \frac{\bar{k}_l(\bar{k}_l+iz_0)}{\Xi_l}\; \Omega_l \; u_l\; 
e^{-i\varphi/2},\\
t^{eh}_{NS}(l) = & \frac{-iz_0 \bar{k}_l}{\Xi_l}\; \Omega_l \; v_l\;  e^{i\varphi/2},\\
\Xi_l = & Ez_0^2+\bar{k}^2_l \left( \frac{E+\Omega_l}{2}\right).
\end{align}
For instance, $t^{he}_{NS}(l)$ is the transmission coefficients from the electron branch in 
a superconductor to the hole branch in a normal metal.
In above coefficients, we use a relation $q^\pm_l=k^\pm_l \simeq k_l$ for simplicity.
The conclusions in this paper remain unchanged in this approximation.

\begin{widetext}
\section{Green functions}
The real space retarded Green function in clean junctions can be calculated by using 
the transmission and the reflection 
coefficients in Appendix A. 
For $ x< x' < 0$, the Green function from a normal metal to a normal metal is
\begin{align}
&\hat{G}_0^{NN}(\boldsymbol{r},\boldsymbol{r}') = -i \frac{\pi N_0}{W} \sum_{k_y^{(l)}}
e^{ik_y^{(l)}(y-y')}
\nonumber \\
\times&\left[ \frac{1}{q^+}\left\{ 
\left(\begin{array}{cc} 1 & 0 \\ 0& 0 \end{array}\right) e^{iq^+_l|x-x'|} + 
\left(\begin{array}{cc} 0 & 0 \\ 1& 0 \end{array}\right) e^{iq^-_l x}e^{-iq^+_l x'} 
\;r^{he}_{NN}(l) +
\left(\begin{array}{cc} 1 & 0 \\ 0 & 0 \end{array}\right) e^{-iq^+_l(x+x')} \;r^{ee}_{NN}(l) 
\right\}\right. \nonumber\\
&+ \left. \frac{1}{q^-}\left\{
\left(\begin{array}{cc} 0 & 0 \\ 0& 1 \end{array}\right) e^{-iq^-_l|x-x'|}
+\left(\begin{array}{cc} 0 & 1 \\ 0& 0 \end{array}\right) e^{-iq^+_lx}e^{iq^-_lx'} 
\;r^{eh}_{NN}(l) 
+\left(\begin{array}{cc} 0 & 0 \\ 0 & 1 \end{array}\right) e^{iq^-_l(x+x')} \;r^{hh}_{NN}(l) 
\right\}\right],\\
& N_0 = \frac{m}{\pi \hbar^2}.
\end{align}
For $ x> x' > 0$, the Green function from a superconductor to a superconductor is
\begin{align}
&\hat{G}_0^{SS}(\boldsymbol{r},\boldsymbol{r}') = -i \frac{\pi N_0}{W} 
\sum_{k_y^{(l)}}e^{ik_y^{(l)}(y-y')}
\frac{E}{\Omega_l} \nonumber \\
\times&\hat{\Phi}\left[ \frac{1}{k^+}\left\{
\left(\begin{array}{cc} u^2_l & u_lv_l \\ u_lv_l& v_l^2 \end{array}\right) 
e^{ik_l^+ |x-x'|} 
+ 
\left(\begin{array}{cc} -u_lv_l & v_l^2 \\ u_l^2& -u_lv_l \end{array}\right) 
e^{-ik^-_lx +ik^+_lx'} \; r^{he}_{SS}(l) 
+\left(\begin{array}{cc} u^2_l & -u_lv_l \\ u_lv_l & -v_l^2 \end{array}\right) 
e^{ik_l^+(x+x')}
\; r^{ee}_{SS}(l) \right\}\right.\nonumber \\ 
&+ \frac{1}{k^-}\left\{ 
\left(\begin{array}{cc} v_l^2 & -u_lv_l \\ -u_lv_l& u_l^2 \end{array}\right) 
e^{-ik_l^-|x-x'|}
+ \left. \left(\begin{array}{cc} u_lv_l & u_l^2 \\ v_l^2& u_lv_l \end{array}\right) 
e^{ik_l^+x -ik^-_l x'} \; r^{eh}_{SS}(l) 
+\left(\begin{array}{cc} -v_l^2 & -u_lv_l \\ u_lv_l & u_l^2 \end{array}\right) 
e^{-ik_l^-(x+x')} 
\; r^{hh}_{SS}(l)\right\}
\right] \hat{\Phi}^\ast,\\
&\hat{\Phi}=\left(\begin{array}{cc} e^{i\frac{\varphi}{2}} & 0 \\ 0 & e^{-i\frac{\varphi}{2}} 
\end{array}\right),
\end{align}
where $\varphi$ is the phase of a superconductor.
For $ x>0> x'$, the Green function from a normal metal to a superconductor is
\begin{align}
\hat{G}_0^{SN}(\boldsymbol{r},\boldsymbol{r}') =& -i \frac{\pi N_0}{W} 
\sum_{k_y^{(l)}}e^{ik_y^{(l)}(y-y')}
  \nonumber \\
\times&\hat{\Phi}\left[ \frac{1}{q^+} \left\{
\left(\begin{array}{cc} u_l & 0 \\ v_l& 0 \end{array}\right) 
e^{ik^+_lx -i q^+_l x'} 
\; t^{ee}_{SN}(l)
+ 
\left(\begin{array}{cc} -v_l & 0 \\ u_l& 0 \end{array}\right) e^{-ik_l^-x -i q_l^+x'} 
\; t^{he}_{SN}(l) \right\}
\right.\nonumber \\ 
&+ \left. \frac{1}{q^-} \left\{ \left(\begin{array}{cc} 0 & -v_l \\ 0& u_l \end{array}\right) 
e^{-ik_l^-x + iq^-x'} \; t^{hh}_{SN}(l) 
+\left(\begin{array}{cc} 0 & u_l \\ 0 & v_l \end{array}\right) 
e^{ik_l^+x +i q^-_lx'} 
\; t^{eh}_{SN}(l) \right\}
\right]. 
\end{align}
For $ x<0< x'$, the Green function from a superconductor to a normal metal is
\begin{align}
\hat{G}_0^{NS}(\boldsymbol{r},\boldsymbol{r}') =& -i \frac{\pi N_0}{W} \sum_{k_y^{(l)}}
e^{ik_y^{(l)}(y-y')}
\frac{E}{\Omega_l} \nonumber \\
\times&\left[ \frac{1}{k^+} \left\{
\left(\begin{array}{cc} u_l & -v_l \\ 0& 0 \end{array}\right) e^{-iq^+_lx  
+ik^+_lx'}
\; t^{ee}_{NS}(l)
+ 
\left(\begin{array}{cc} 0 & 0 \\ u_l& -v_l \end{array}\right) 
e^{iq-_lx+ik^+_lx'} 
\; t^{he}_{NS}(l) \right\}
\right.\nonumber \\ 
&+ \left. \frac{1}{k^-} \left\{ \left(\begin{array}{cc} 0 & 0 \\ v_l& u_l \end{array}\right) 
 e^{iq^-_lx-ik^-_lx'} \; t^{hh}_{NS}(l) 
+\left(\begin{array}{cc} v_l & u_l \\ 0 & 0 \end{array}\right) e^{-iq^+_lx 
-ik^-_lx'} 
\; t^{eh}_{NS} (l)\right\}
\right] \hat{\Phi}^\ast. 
\end{align}
\end{widetext}

\end{document}